# The Entropy of Attention and Popularity in YouTube Videos

Jonathan Scott Morgan
Michigan State University School of Journalism – morga125@msu.edu

Iman Barjasteh
Michigan State University College of Engineering – barjaste@msu.edu

Cliff Lampe
University of Michigan School of Information – cacl@umich.edu

Hayder Radha
Michigan State University College of Engineering – radha@egr.msu.edu

**Abstract**
The vast majority of YouTube videos never become popular, languishing in obscurity with few views, no likes, and no comments. We use information theoretical measures based on entropy to examine how time series distributions of common measures of popularity in videos from YouTube's "Trending videos" and "Most recent" video feeds relate to the theoretical concept of attention. While most of the videos in the "Most recent" feed are never popular, some 20% of them have distributions of attention metrics and measures of entropy that are similar to distributions for "Trending videos". We analyze how the 20% of "Most recent" videos that become somewhat popular differ from the 80% that do not, then compare these popular "Most recent" videos to different subsets of "Trending videos" to try to characterize and compare the attention each receives.

**Acknowledgements**: This work has been supported by the National Science Foundation under Grant IIS-0968495.

## 1    Introduction

There is a tension between the amount of content posted to social media sites and the attention people using those sites can spend on consuming content. These sites change rapidly as new material is added, and attention can easily wander from post to post. This ease of attention shifting may have implications for our ability to bring sustained attention to important issues, or design systems that help people manage their attention. In this study, we use information theoretical measures of entropy and information divergence to examine how attention given to YouTube videos, measured by views, comments and likes, forms and changes over time, particularly near the beginning of a video's life.

Attention paid to content in social media channels can be powerful, but fickle. Video and photos of police pointing sniper rifles and assault weapons at and threatening protestors in the aftermath of the shooting of Michael Brown in Ferguson, MO, inspired a national discussion of militarization of the police (Bosman & Apuzzo, 2014; Golgowski, Wagner, & Siemaszko, 2014). The YouTube video of Karen Klein being harassed by students inspired media attention on bullying and over $700,000 in Internet donations to Klein (Preston, 2012; Thomas, 2012), but also inspired death threats and threatening emails, online comments, and calls directed at the bullies (Goldman, 2012). A YouTube video about the atrocities of Joseph Kony, leader of the Lord's Resistance Army in Uganda, garnered more than 86 million views in three weeks and inspired mass media stories, but also opened critiques of the video and its makers that caused one of that organization's leaders to have a psychotic episode (Rainie, Hitlin, Jurkowitz, Dimock, & Neidorf, 2012). Viral video from the Middle East played a substantial part in inspiring and galvanizing recent Arab Spring uprisings (Anderson, n.d.; Tufekci & Wilson, 2012), but social media have also helped terrorist groups like ISIS gain power and influence in the chaos that followed (Berger, 2014; Kingsley, 2014; Speri, 2014). And the viral spread of videos of the ALS ice bucket challenge helped the ALS Association raise $100 million dollars in a span where it raised $2.8 million the previous year, but wasn't

initiated by the organization and isn't easily reproducible, highlighting the fickle and unpredictable nature of viral fundraising (MacAskill, 2014; Pallotta, 2014).

These examples show the substantial potential for shared tweets, images, and videos to inspire large-scale collective action. They also show that the resulting complex individual and shared reactions make it difficult to harness or predict the effects of this collective attention. This kind of mass public attention is the exception, however, not the rule. Both individual and collective attention are scarce resources, and so most content on the Internet neither captures much individual attention or inspires action.

In this study, we explore how measures of individual attention paid to YouTube videos (views, likes, and comments) help to characterize and differentiate between types of attention within YouTube.

Past studies of attention in YouTube have found a roughly linear relationship between log-transformed counts of views from relatively early on in a video's life cycle and counts from 5, 7, and even 90 days later (Borghol, Ardon, Carlsson, Eager, & Mahanti, 2012; Cha, Kwak, Rodriguez, Ahn, & Moon, 2007; Cheng, Dale, & Liu, 2008). While videos can have spikes later in life based on external links (Crane & Sornette, 2008), most videos have a relatively smooth line of accruing views over their active life-cycle (which varies in duration depending on a number of factors), then their viewership decays substantially.

Researchers disagree on how best to model this growth and subsequent decay, but most agree that the growth of videos is, in general, relatively stable and linear, except at the beginning of a video's life. Most of the research to date hasn't had fine-grained enough data to explore this period, and so there is relatively little insight into how the relatively stable long-run rate of a video's growth is set in the period just after its upload.

To examine this period of flux in more detail, we gathered viewing data every 6 hours from videos as they were posted to the "Recently Posted" and "Trending" video feeds on YouTube. We segmented the videos into a number of different categories and examined entropy and information divergence to see how metrics that represent attention behave and relate to each other over the videos' effective lifetime. Being the best-known measure of uncertainty, entropy allows us to look for periods in the life cycle of a video where attention is fluid (uncertain), and so helps us pinpoint periods where a video's attention trajectory is more likely able to change (perhaps toward a more certain level of attention). Information divergence allows us to characterize and attempt to differentiate between the distributions of attention of different sets of videos, broken out by levels of popularity. The combination of statistics lets us create a more detailed profile of the life of a video – when changes in attention can alter the trajectory of a video's popularity, and when that trajectory becomes mostly stable.

## 2    Literature Review

Attention at both the individual and public levels has been studied for decades. Researchers in diverse fields have focused on both the broader public's attention and the idea of individual attention as an information consumption capacity made scarce by an individual's limited time and cognitive resources, pioneered by Herbert Simon and Michael Goldhaber. As access to the Internet and use of social sharing systems increase, researchers have migrated these concepts of attention from the realm of mass media to online social sharing systems, where the amount of information available to and targeted at an individual is increasing quickly, making attention ever more precious.

### 2.1    Individual Attention

Attention is defined by Herbert Simon as an individual's capacity for consuming information, limited by the time and cognitive resources they can devote to consumption (Simon, 1971). This concept is based on Simon's "bounded rationality": individuals have cognitive limitations that keep them from behaving rationally (March & Simon, 1958), especially in situations where the complexity of the environment is immensely greater than their cognitive ability to deal with that complexity (Simon, 1996). This leads to an attention economy defined by Simon as having the following traits: 1) there's a finite limit to attention; 2) people tend to conserve cognitive resources; 3) it's costly to decide how to spend attention and switch it once an initial choice has been made; and 4) people are using some mechanism to satisfice their attention choices.

In this attention economy, limited time and cognitive resources along with the human brain's inability to make decisions in parallel (Levy, Pashler, & Boer, 2006; Smith, 1967) make attention a scarce resource, one that becomes increasingly valuable as the Internet and social network and social media systems provide more and more channels and information items from which an information consumer must choose (Goldhaber, 1997; Simon, 1996).

## 2.2  Attention and Social Media

As more people gain access to the Internet and begin to use social media sites like Facebook, Google+, Twitter, YouTube, Flickr, Instagram and Pinterest, there's a complex design interplay between the need to provide access to diverse and plentiful information, while still providing tools that help users manage how much information they are exposed to. As Simon (1996) says (Simon, 1996): "The task is not to design information-distributing systems but intelligent information-filtering systems. (p. 144)".

Various definitions of attention provide insight into how people choose which channels to engage with, how much time to spend on each, and what to interact with in a given channel. Attention paid to "memes" on the Internet tends to shift quickly in social media environments (Leskovec, Backstrom, & Kleinberg, 2009; Simmons, Adamic, & Adar, 2011).  In the context of seeking users' attention for new sites or mediums, Gilbert et al. found that it is difficult to get users to switch to new tools, even when there was a clear benefit to be gained from switching (Gilbert, Karahalios, & Sandvig, 2008).  In Facebook, researchers defined attention as carrying out one of a range of actions that require different levels of effort (from viewing profiles or pictures to writing comments, wall posts, and private messages) and found that users not only pick and choose whom they interact with, but also vary the level of effort they put into interaction based on the person with whom they are interacting (Backstrom, Bakshy, Kleinberg, Lento, & Rosenn, 2011).

There are many theories that could potentially explain how people make the types of attention decisions outlined above.  Social cognitive theory and social learning (R. LaRose, Mastro, & Eastin, 2001), selective exposure (Baran & Davis, 2008; Sears & Preedman, 1967), uses and gratifications theory (Joinson, 2008; Lampe, Wash, Velasquez, & Ozkaya, 2010; Papacharissi & Mendelson, 2011) and habit (Robert LaRose, 2010; Wohn, Velasquez, Bjornrud, & Lampe, 2012) are all thought to play a role in how individuals spend their attention.  Given the complexity of the decisions being made, no one of these theories is likely able to explain all decisions.  How people direct attention in social media is likely a complex interplay between these theories, tasks dependencies, and the features of each site to be potentially used.

## 2.3  Attention and YouTube

A study of the effects of geography and localization of content on video popularity showed that despite YouTube being a global site, the popularity of YouTube videos is constrained by the geographic locality of a given video's topic (Brodersen, Scellato, & Wattenhofer, 2012).  Studies of YouTube's role in fostering political discourse and marshaling public attention found that users engage strongly with political content on YouTube, indicating the potential for it to grow in political influence as it grows in popularity (Garcia, Mendez, Serdült, & Schweitzer, 2012; Gueorguieva, 2008).

Traits of the creator impact attention received.  Researchers have found that an uploader having more social ties (Borghol et al., 2012; Rodrigues, Benevenuto, Almeida, Almeida, & Gonçalves, 2010) and focusing on certain groups of connections (Spathis & Gorcitz, 2011) increases video popularity.

References to videos also figure prominently in the attention a video receives.  Crane shows how external links can increase a video's views substantially if they come at the right time (Crane & Sornette, 2008).  Zhou et al. found that referrals from YouTube's "Related Videos" list are, for most videos, a larger source of views than search referrals (Zhou, Khemmarat, & Gao, 2010).

## 2.4  Predicting Attention in YouTube

Varied strategies have been used to predict future YouTube views based on past attention data, including stochastic models (Mathioudakis, Koudas, & Marbach, 2010), neural network-based reservoir computing (Wu, Timmers, Vleeschauwer, & Leekwijck, 2010), genetic algorithms (Kender, Hill, Natsev, Smith, & Xie, 2010), and a series of studies based on statistical models.

At a high level, researchers have found that log-transformed YouTube views generally have a power law distribution with exponential decay (a truncated tail).  Cha et al. in 2007 (Cha et al., 2007) and Szabo and Huberman in 2010 (Szabo & Huberman, 2010) used daily counts of views to find and support that pairs of log-transformed daily view counts across the life of a video correlate highly (Cha found r = .84 between views on $2^{nd}$ day and $90^{th}$ day and Szabo found r = .92 between $7^{th}$ day and $30^{th}$ day) and have a log-linear relationship.  Crane and Sornette broke out types of attention and showed how this common distribution would be disrupted, in certain circumstances, when external links suddenly focus attention on particular videos (Crane & Sornette, 2008).  Using weekly counts of views, Cheng et al. also found that attention was a function of the video's rate of attention change and time, independent of initial views (Cheng et al., 2008).

Researchers have also worked to create models that more precisely capture the nature of attention's eventual decay. Avramova et al. implemented a non-linear model that decides between exponential and power law decay based on the shape of a video's view distribution (Avramova, Wittevrongel, Bruneel, & De Vleeschauwer, 2009). Borghol et al. made a multi-linear model that included traits of the video with the previous time period's views to predict a given time period's views (Borghol et al., 2012). They also found that early view counts didn't correlate as well with future views as view counts collected later in the video's life. Pinto et al. explored the independence of attention from initial views by predicting future attention using daily view data in a dynamic linear regression model that could target any day in a video's life (Pinto, Almeida, & Goncalves, 2013). To account for different rates of initial attention, it included each of the previous days' view counts, weighted either based on that day's proportion of overall viewing or based on how the overall distribution compared to a set of patterns the researchers had established.

## 3 METHODS

### 3.1 Operationalizing Attention in YouTube

Videos and video channels are the main information units in YouTube. A registered member of YouTube can follow another user's video channels, but cannot follow other users. Attention in YouTube is also video-centric. Videos can be watched, "like"d, commented on, and shared; and only the views, likes, and comments are tracked on the site. This allows for only a few basic measures of attention: views, likes, and comments.

### 3.2 Data Collection

To study the life cycle of videos on YouTube, we collected newly uploaded videos from the only two standard feeds in YouTube's Data API that capture recently uploaded videos: the "Trending videos" feed, subsequently referred to as "trending", and the "Most recent" feed, subsequently referred to as "recent". Each feed returned 150 total videos each time it was checked, and videos only appeared in one or the other of the feeds, never both. Each time we checked a feed, we compared the videos returned to the database of videos that we tracked for this study. When a video was in a feed and not present in the database, we added it. There were always at least some videos in the feed that had also been there in our previous check, indicating that a six-hour interval was sufficiently frequent to get all the videos added to each list.

We checked each of these two feeds every six hours, averaging 24 new trending videos per day and 69 new recent videos per day. We collected fine-grained attention-related data on a total of 1,460 trending and 4,250 recent videos that we identified as being unique based on monitoring the two feeds (trending and recent) over a two month period from 9/21/2012 to 11/27/2012, then we monitored these videos' statistics for more than a year to ensure the validity of our analysis. Network and hardware problems occasionally interrupted collection, causing the total number of videos collected to be a bit lower than the average would predict. Of the videos collected, 1,002 trending and 2,323 recent videos stayed active (were not taken down for whatever reason) for at least two weeks, and so were included in our study.

To capture change over time, every six hours a separate process looped over all of the videos in our database and retrieved updated meta-data for them, including views, likes, and comment count. We chose a six hour interval for fetching updated attention data for two reasons: 1) YouTube doesn't update the meta-data for its videos in real time. It updates them at an interval that depends on how busy their servers are, anywhere from every 30 minutes to every 2 hours, and sometimes even less frequently. Our tests of different time intervals indicated that collecting attention data every 6 hours left enough time in between collections that most videos' information had been updated since the previous collection; 2) In addition, we chose an interval that left us with enough hardware and network bandwidth that we could reliably and consistently continue to update this information as our data set grew. We aimed to collect two weeks of data on each video.

There were a few instances during our data collection when network or hardware problems caused us to miss am update collection interval. In those instances, we estimated the data for the missed interval (views, comments, and likes) by averaging the values of each score from the interval before and after.

## 3.3 Correlation Between Attention Measures

We first looked at correlations between the three available attention metrics (views, likes, and comments) over the life of each video to assess if the metrics were likely all proxies for a common trait of attention. To do this, we created a time series of each of the three attention metrics for each video in our data set, then for each video we did pair-wise correlations of each of attention time-series with the others, resulting in three correlation scores per video. We then used a histogram of the distributions for each of these three scores to assess the general levels of correlation between the attention traits across videos within trending and recent videos (Figures 1 and 2).

## 3.4 Entropy and Information Divergence

To more closely examine attention paid to videos, we employed two information-theoretic measures to quantify and analyze different statistical aspects of attention. First, we used the Shannon entropy measure $H$ to quantify the level of randomness in attention distributions and quantify and highlight how different attention distributions vary over time. Second, we used a normalized information-divergence measure $\Delta$ to quantify how different the statistical distributions of attention can be from each other for any pair of groups of videos (e.g., trending versus non-trending videos). We chose these information theoretical measures because we are mainly interested in comparing the *distributions* of attention that videos of different popularity levels receive over time, independent of the magnitude of the attention a given video receives.

YouTube videos can receive vastly different magnitudes of attention. In the data collected for our study, for example: among trending videos the maximum total views for a video was 69,382,458; the mean was 1,134,000; the median was 117,398; and the mode was 18,008. Among recently added videos the maximum views for a video was 65,509; the mean was 692.13; the median was 212; and the mode was 15. The range of total views within trending videos is wide, and centrality measures show that while there are substantial outliers (mean of 1,134,000 is far larger than median 117,398), there are also substantial differences in total views across the set of trending videos. Recently added videos have a much more narrow range, but it doesn't matter since they are included in comparisons with the trending videos.

In order to analyze and compare the distributions of attention metrics for videos with this wide a range of attention, we need statistics that are robust when used to compare distributions that contain values of vastly different magnitudes. Traditional statistics that compare distributions using the actual values in a set of numbers are sensitive to large differences in values, and so would be misleading if used to analyze this data given the wide range of magnitudes of attention, in particular in views. The Entropy and Information Divergence measures used in our study are based on probability distributions that characterize each distribution independent of the particular values it contains, allowing for analysis and comparison even when the sets of values are of substantially different magnitudes.

In our study, both Entropy and Information Divergence depend on the probability distribution of a given attention metric in a group of videos at a certain point in time $p(t)$. This distribution will show the range of attention being given to videos in that group at that time – it could have a high degree of variation if different videos in the group are getting different amounts of attention, or it could be uniform if videos in the group are getting similar levels of attention. To calculate entropy and information divergence in this study, we first calculate the probability distribution of attention for a set of videos at each point in time in our time series. We then use these probability distributions to calculate the entropy and information divergence scores used to capture and analyze changes in attention over time in our analysis.

Entropy captures the level of randomness within the values in a given distribution $p(t)$. It is calculated using:

$$H\big(p(t)\big) = \sum_i p_i(t) \log(1/p_i(t))$$

where $H$ is greater than or equal to 0 and 0 denotes no entropy or randomness. If videos are receiving similar levels of attention in a given distribution, then there is some consistency in the attention the videos are receiving, and so randomness and entropy will be low. Consequently, when $H$ is close to zero, attention will be virtually deterministic, and hence, it will be easier to predict for a given group of videos that exhibit the underlying distribution $p(t)$. If videos in a group are receiving wildly different amounts of attention (some low, some high, and some in between), this attention is inconsistent and indicates some randomness, and so entropy would be high. In our study, we examine how entropy of attention in groups of videos changes over time to assess where in videos' overall life cycle there is more potential for change in the trajectory of attention.

Normalized information divergence (∆) quantifies differences between two distributions. For two distributions $p$ and $q$, Shannon information divergence is defined as follows:

$$D(p||q) = \sum_i p_i(t) \log(p_i(t)/q_i(t))$$

This definition for information divergence is asymmetric - $D(p||q) \neq D(q||p)$. Using this standard Shannon information divergence $D(p||q)$, however, one can define a symmetric divergence measure $\Delta(p,q)$ such that $\Delta(p,q) = \Delta(q,p)$:

$$\Delta(p,q) = \frac{D(p||q) + D(q||p)}{H(p) + H(q)}$$

To simplify our analysis, we use the second expression for information divergence, ∆, to create a symmetric, normalized information divergence score. In the formula for this score, ∆, normalizing by the entropies $H(p)$ and $H(q)$ of the two distributions provides an explicit measure of the level of information divergence relative to the level of uncertainty (entropy) associated with the two distributions $p$ and $q$. It is important to note that information divergence in general, and hence the measure ∆, is always nonnegative. Consequently, we have $\Delta \geq 0$, where 0 indicates no divergence between the two distributions (i.e., the two distributions are identical). If two groups of videos have substantially different levels of attention across a time period, the information divergence score comparing the two would be high – perhaps 1, or even as high as 2 if there are spikes in attention. If a part of the distribution differs but they converge as time goes on, ∆ would likely be a modest 0.3 or 0.4. If the two distributions are very close throughout the time period, then ∆ will likely be small: 0.1 or less. In addition to measuring differences in time-series distributions, ∆ can also be used to assess changes in data over time – distributions can be made at different time intervals, then ∆ can be calculated between contiguous pairs.

## 3.5 Comparing Attention

The videos in our sample had vastly different magnitudes of viewership. To make all of the videos in our data set comparable, we normalized the time-series distributions of 2 weeks of attention for each video so each time period's value is the proportion of the overall attention represented by the count in that time slice, rather than making each distribution from simple counts of attention measures. All time series data we derived was created at the highest resolution we have for changes to videos – increments of six hours.

We created two different normalized distributions for each video, one where attention was aggregated cumulatively over time (used in our analysis of information divergence, to implement the theoretical idea that previous attention affects future attention in social media sites), and one where only the attention during that time period was captured, not including any attention previously accrued (used in our analysis of entropy, so we specifically targeting the randomness in each time period, without potential interference from including popularity at previous times). We then calculated the distributions of attention at each time point for each group, and calculated ∆ between these individual time period distributions.

We chose to focus on views in our comparisons of different groupings of videos. Comments and likes are present in many videos, but not uniformly across all trending and recently added videos (95.41% of trending videos had comments, for example, but only 15.69% of recently added videos did), and in much smaller numbers on average than views (in trending videos, maximum comment count was 308,484, mean was 2877.7, and median was 212; while in recently added videos, maximum comment count was 174, mean was 4.39, and median was 2. Likes were even less prevalent, with many videos not having any at all.), creating potential statistical problems because of lack of variance, in particular among recently added videos. Since the correlations between views, likes, and comments described in the next section suggest that all three metrics are relatively good proxies for attention, we decided to focus on the measure that had the most data associated with it, views.

# 4 RESULTS

## 4.1 Correlation Distributions

Our first attempt to characterize differences between recent and trending videos was to correlate all possible pairs of the time series of each video's views, likes, and comments over the first 14 days of the videos life, then make histograms of the distributions of those correlations in trending (Figure 1) and recent (Figure 2) videos, to see how these three traits that represent attention related in the two broad groups – videos.

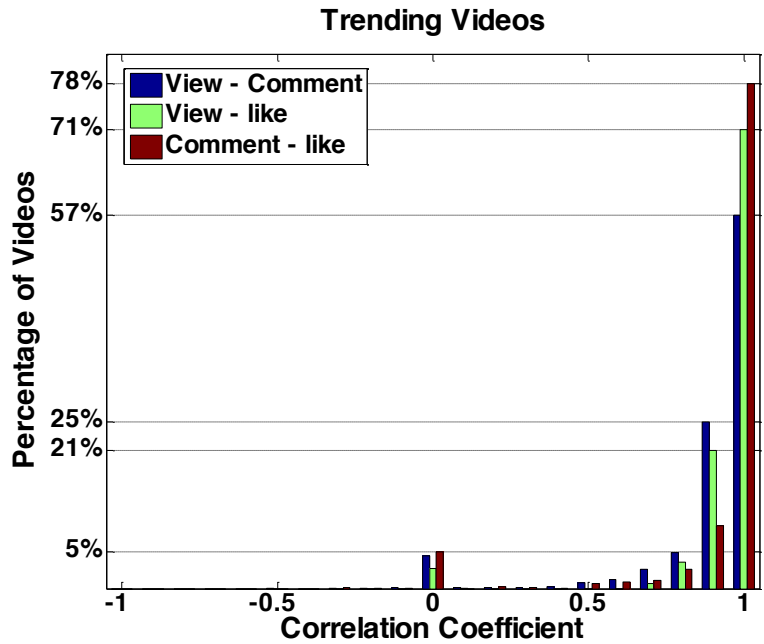

Figure 1. Distribution of correlations between attention traits for trending videos.

The correlation histogram for trending videos (Figure 1) shows that views, comments and likes are highly correlated with each other and with overall popularity.  Most trending videos eventually become popular, and in most trending videos (86% or more of the total number of videos), all three traits correlate strongly with each other, where the correlation coefficient $\rho$ is between 0.8 and 1.  The vast majority of most-recently added videos never achieve any significant level of popularity, and so there is little in the time series to correlate; and, hence, correlations are almost all near 0 (Figure 2).

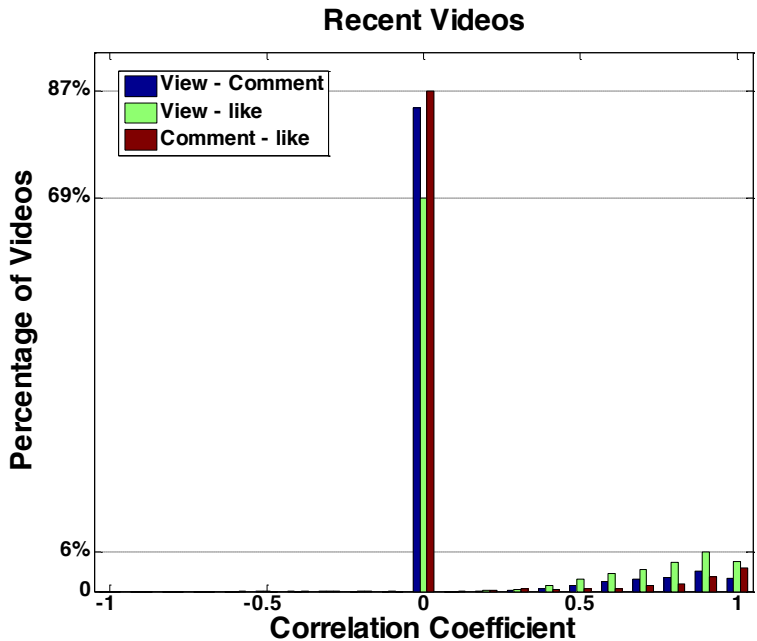

Figure 2. Distribution of correlations between attention traits for recently added videos.

In each case, the correlations behave very similarly depending on the level of popularity of the video.  Attention metrics for recent videos tend to all not correlate, and those same metrics for trending videos seem to all correlate, and at roughly the same level.  This suggests that these three traits are reflecting the broader popularity of the video, not different facets of popularity.

The correlation-coefficient histogram for recent videos (Figure 2) also helps us to better characterize popularity in YouTube – a relatively small percentage of recently added videos have positive correlations that are near 1, just like trending videos. This suggests that some of these recently added videos do achieve a certain level of popularity, and this led us to explore in more depth the recent videos that did capture some attention.

To compare the most popular recent videos to the videos in the trending feed, we first made a grouping of the 20% of our sample of recent (R) videos where views, comments and likes were present enough to correlate relatively highly (referred to subsequently as R5). We then broke our trending video sample into 5 equal quintiles as well, sorted by increasing total numbers of views received. These five groupings of trending (T) videos are referred to as T1 through T5, from lowest numbers of views to highest numbers of views, where T1 is the 20% of our trending video sample with the lowest number of views and T5 is the 20% of our trending video sample with the highest numbers of views. No videos appeared in both the recently added and trending feeds, so there is no overlap between R5 and the T groupings.

## 4.2  Entropy

Entropy measures the amount of inconsistency or randomness in a distribution of numbers. In the context of YouTube videos, we measured entropy to assess where in videos' overall life cycles attention is random or inconsistent, and so where there is potential for change in the trajectory of attention. For each of our six basic groupings of videos (top 20% of recent videos, R5, and then 5 equal 20% groups of trending videos broken out based on increasing attention, from T1 to T5), we calculated and plotted the entropy statistic for each of the six-hour time-periods in our time series (Figure 3). We then calculated the mean and variance of the six entropy values (one each for R5, T1, T2, T3, T4, and T5) for each time period and created a dot-and-whisker plot to give a clearer picture of the range of entropy at each time period (Figure 4).

### Entropy of R5, T1, T2, T3, T4 and T5

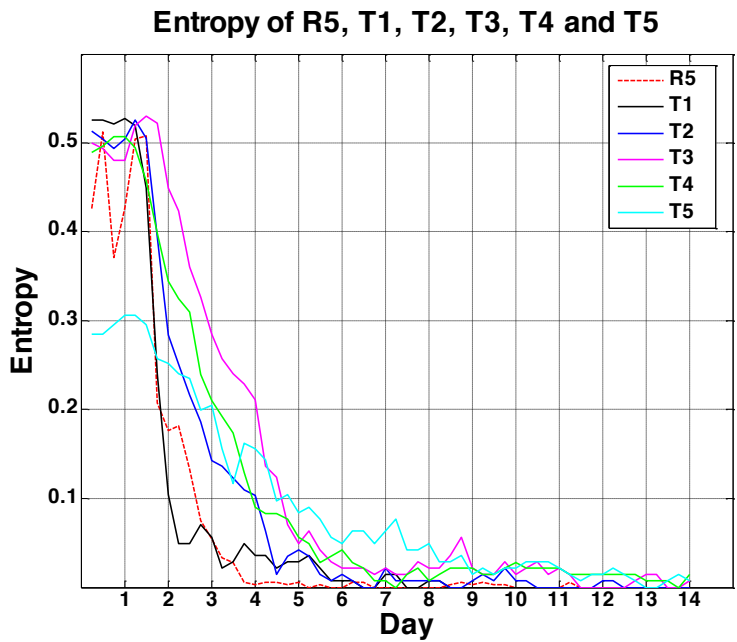

Figure 3. Distributions of entropy over time for each of our 6 basic groupings of videos.

The raw graph of the entropy values (Figure 3) looks a little chaotic, but there is a clear trend of entropy being at its highest the first 2 days of a video's life (between 0.3 and 0.5, with most around 0.5), then decreasing rapidly toward 0 over the 3rd through 5th days, and remaining relatively stable at close to 0.1 from then on. The most popular 20% of videos in trending maintain a small amount of entropy longer than the others, and there are small spikes, but after the 5th day, entropy stabilizes under 0.1, an entropy value that represents a very small amount of randomness.

**Mean and Variance of Entropy of R5, T1, T2, T3, T4, T5**

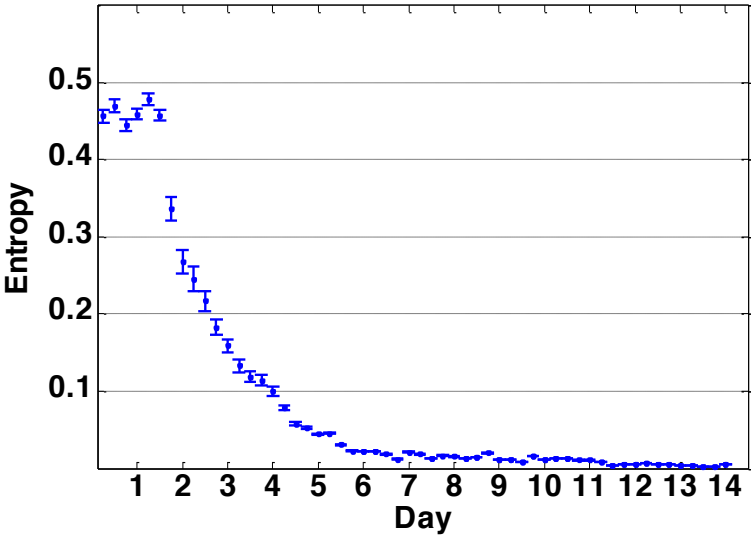

Figure 4. dot-and-whisker plot of mean and variance of entropy, by 6-hour time period, where each "x" plots mean entropy and bars above and below show variance.

Plotting the mean and variance of each (Figure 4) makes the consistency of this trend across our video groupings more clear. Average entropy starts out high, about 0.45, with relatively small variance – all videos have pretty much the same level of variance early on. Then, as entropy decreases, the variance does increase some, but only for three or four time periods before it returns to the modest levels of variance seen at the top. Then, as entropy decreases, the variance narrows even further as the entropy approaches zero on around the 5[th] day – at around the 4[th] day, entropy converges across all videos and decreases uniformly.

## 4.3   Information Divergence

Information divergence quantifies the difference between two distributions. We use information divergence to look at differences between attention paid to videos in pairs of our quintile groups over time, and to look at how much attention changes from time period to time period within these groups of videos.

Both of these comparisons calculate divergence using a set of distributions calculated per group for 14 days worth of 6-hour time periods. For each time period, we measured the cumulative attention paid to each video up to that time-period. This attention measure represents the total viewership that each video achieves over its entire life up to that point in time; and it is the same number one can record or observe when accessing the YouTube API.

### 4.3.1   Comparing Attention Between Groups

To compare attention distribution between groups, for each pair of groups of videos from R5, T1, T2, T3, T4, and T5, we iterated over the 14 days worth of 6-hour time periods, creating probability distributions of attention for each grouping, then calculating the divergence between each pair of groups' attention distributions for each time period. We plotted these pair-wise divergence results as functions of time, side-by-side for comparison (Figure 5).

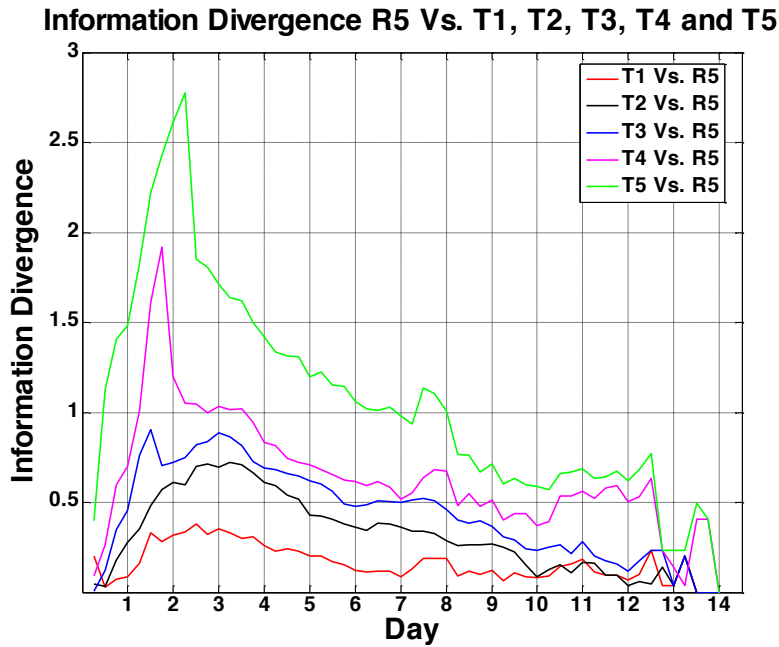

Figure 5. Information divergence comparing attention distributions between groups.

Figure 5 shows the difference in attention (distribution of views) over the first 14 days of a video's life cycle between the top 20% of recent videos (R5) and all five quintiles of trending videos (T1 through T5). In terms of R5, this plot indicates that the attention distribution of R5 is very similar to that of T1 - the average divergence is 0.1624, which is relatively small. Thus, the attention profiles of the least-popular trending videos, T1, and the most popular recently-uploaded, R5, are very similar. The attention profiles of the other, more popular quintiles of trending videos are increasingly different, and the divergence becomes more and more concentrated on the left edge of the distribution, where our entropy results suggest changes in attention decide the attention trajectory of videos. This is especially evident when comparing R5 with T4 and R5 with T5. There is a huge divergence (greater than 2) over the first three days that suggests that these very popular videos have substantially different attention patterns from R5 (and so likely from T1 and T2, as well).

We also compared the different quintiles of trending videos with each other, and found that contiguous quintiles consistently have very low divergence (between 0.0752 and 0.1213), and that the divergence increases the larger the distance between quartiles (distance of two is between 0.1647 and 0.2433; distance of three had divergence of 0.2729, from 1 to 4, and 0.5056, from 2 to 5; and distance of four, from 1 to 5, had divergence of 0.5634). The attention distributions for the trending quintiles aren't that different from those next to them, but they become more different as the two quintiles are farther apart.

### 4.3.2    Examining Attention Change Within Groups

To compare attention within our groups of videos over time, we calculated the divergence between the attention distributions in every contiguous pair of time periods in our time series data and plotted the divergence scores (Figure 6).

**Time-Lagged Information Divergence of R5, T1, T2, T3, T4 and T5**

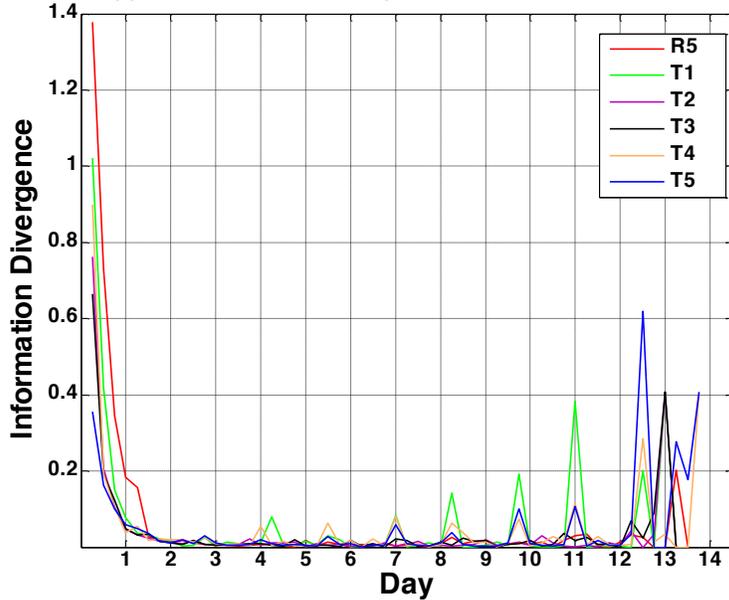

Figure 6. Information divergence comparison of attention change between contiguous pairs of time-lagged time periods.

Figure 6 shows how attention changes between contiguous pairs of time-lagged time periods. This figure indicates that the most substantial within-group changes in attention occur within the first day of a video's life. The magnitude of this volatility is different between groups, but for each their largest changes in attention profile are at the start of their life cycle. The trending videos do have some subsequent volatility, likely because they are featured more prominently, and so have more chances to be discovered later in their life cycle.

## 5   DISCUSSION

We are interested in how attention paid to social media content changes over time, and when and how quickly those changes tend to occur. In YouTube in particular, where past attention has a reliable log-linear relationship with future attention after the 3$^{rd}$ day or so of a video's life, we are interested in better understanding how and when the slope of this attention trajectory is set early in a video's life.

Entropy was a substantial help in quantifying periods where change in attention was more likely. Analysis of entropy in attention profiles of popular videos indicates that much of the randomness or uncertainty in a video's lifecycle occurs within the first two days of the video being posted. After that point, there is entropy, but it decreases quickly within a few days and becomes uniformly very low (less than 0.1), indicating that the distribution of attention for videos in these groups stabilizes and becomes much more predictable after a few days (matching previous research).

Comparing information divergence of attention profiles between groups of videos shows that the popular recently added videos (R5) have an attention profile very similar to the least popular quintile of trending videos (T1), somewhat similar to the second quintile (T2), and very different from the other quintiles (T3, T4, T5), particularly in the early days of videos' life cycle, where the divergence spikes above two when the popular recent videos (R5) are compared to the two most popular quintiles of trending videos (T4 and T5). The relationship between R5, the most popular of the recently added videos, and the trending video quintiles is also much different than that of the trending quintiles with each other. While direct comparison between R5 and T1 suggest that both get similar patterns of attention in terms of their information divergence, R5 and T1 have very different information divergence relationships with T2, T3, T4, and T5 (T1's attention distributions are much more similar to T2, T3, T4, and T5 than R1's are). This suggests that while videos not placed in the "trending" feed can receive attention comparable to some less popular trending videos, there are still underlying differences in the attention each receives not captured in our data.

These differences might be more accurately described and understood with a more nuanced clustering of these videos, but even our arbitrary binning into quintiles of increasing attention reveals meaningful differences. If all videos had similar distributions of attention, you'd expect all pairs of quintiles to have similar or the same levels of information divergence, and for that information divergence to be very close to 0. This is not the case. There are meaningful differences in the attention distributions as videos get more views evidenced even in arbitrary quintiles.

Past research suggests these differences are also not due to a video being present in the "trending" feed. Zhou et al.'s analysis of the sources of views for YouTube videos from 2010 suggests that being declared "trending" isn't neither a substantial direct or indirect driver of attention in and of itself (Zhou et al., 2010). From analysis of sources of traffic for 700,661 videos, Zhou et al. found in that roughly 30% of a video's traffic comes from YouTube search, 30% comes from Related videos placed next to a given video, 13% comes from external or "viral" links, and the rest comes from a series of minor sources including external search engine links, embeds of videos, channel pages, and the YouTube home page. The recently added and trending feeds are not explicitly placed on the YouTube home page, presence in these feeds is not thought to be factored into either search or featured selection algorithms, and none of the rest of these sources of attention are likely consistently influenced by a video's presence in the "trending" or "recently added" feeds, either.

The methods used here combine to provide a more detailed look at underlying forces at work in the early life of a video than either does alone. Early on, entropy analysis indicates that attention is fluid in all videos (there is substantial entropy in video views), even those that are less popular. Analysis of divergence in attention profiles shows that differences in attention distributions are magnified in the initial days after a video is uploaded, and it is here in particular where the videos that end up being really popular have a substantially different attention profile from those that are less popular. After a few days, though, the entropy in attention for all videos essentially goes away, and so while some trending videos have some attention divergence, for the most part, analysis of the divergence between time periods indicates that there is little change in a video's attention distribution after the end of the period where there is substantial entropy for all videos. There is substantial divergence in the first days of video life cycle for all groups, but the trending videos have pockets of variability throughout their life cycle, while the popular recently added videos have little divergence after the first few days. These behaviors provide further evidence of a more complex function of attention at work than a simple initial increase that decides the video's attention trajectory going forward. Trending videos have some capacity to alter their attention profiles throughout their life cycle, while the popular recently added videos seem to be more settled into their attention profiles after the first few days.

These findings provide more context for previous studies' analysis of the usefulness of initial view counts in predicting subsequent attention, and suggest strategies for effectively using early attention data to predict future attention. Previous studies found that long-run attention increased at a consistent rate independent from early viewership data (Borghol et al., 2012; Cheng et al., 2008). This makes sense in the context of the early days of a video's existence being high in entropy – for a simple linear model to be accurate, you'd have to wait until entropy dies down and the attention trend has stabilized. For most videos, attention starts to stabilize after 2 or 3 days. Measuring the entropy within attention for a given set of videos also gives an idea of where you might start to look for the trajectory of attention to be set, and how much noise you are looking at if you try to predict earlier.

Information divergence gives guidance on how to group videos when looking for entropy, and when one might be able to start looking for long-run trends, even amidst entropy. Figure 6 in particular illustrates well that, early on, the information contained in attention distributions of videos with similar magnitudes of traffic will start to converge after a day, much sooner than entropy recedes to effectively 0. With finer-grained data that starts close to when a video is uploaded, one might be able to run a clustering algorithm on the attention features extracted from each of the groups that represent different tiers and kinds of attention (R1, R2,…,T1, T2, …) to train a classifier to examine features of a given video and slot it into one of these potential ranges of attention. Since information divergence appears to converge before entropy decreases, thus might allow you to estimate future attention even while entropy is relatively high.

Our work has limitations. It would be good to explore entropy and information divergence with a larger data set. We were interesting in breaking out videos by type, but we focused first on refining our analysis. While our method allows us to see how entropy changes over the life cycle of a video, it doesn't provide insight into why that entropy can be different from one video to another. For example, some videos have periodic intermittent bursts of popularity that allow them to remain popular for a long time, essentially acting as outliers in this entropic model. What causes the uncertainty in these videos' attention to persist could be explored in future work. We also do not take into account the effects of external links to videos (for example, sharing on Facebook or Twitter), which can have a substantial effect on attention (Crane & Sornette, 2008). And, while our data collection covered a significant period of time, attention distributions that include a longer period of time might provide additional insight.

## 6    CONCLUSION

The social media environment is characterized by a massive amount of content that is rapidly refreshed with new posts. This creates a tension between the available amount of attention and the sheer volume of content that could potentially be viewed in those channels. Being able to detect and predict the potential of a given piece of content for attention could be important in a range of domains of study, from political discourse to marketing and advertising to planning IT infrastructure for serving multimedia files to designing systems that encourage users to produce content. The information theoretical methods used here to examine attention enable comparisons between pieces of content with vastly different levels of attention, and they also allow us to start to more precisely detect and characterize the period in a YouTube video's life where patterns of attention move from uncertainty to stability. These measures, combined with more precise data, provide a more detailed view of how and when a given video's potential for attention is fluid and subsequently solidifies in YouTube, and they have the potential to improve the study of other social media channels as well.

## Table of Figures